\documentclass[
journal=jpclcd,
manuscript=letter,
layout=twocolumn]{achemso}

\usepackage{graphicx}
\usepackage{amsthm}
\usepackage{amsmath}
\usepackage{amssymb}
\usepackage{bm} 
\usepackage{physics}
\usepackage{color}
\usepackage{hyperref}
\usepackage[normalem]{ulem}
\usepackage[english]{babel}
\usepackage{lipsum}  

\usepackage{dsfont}
\usepackage[percent]{overpic}
\usepackage{todonotes}
\usepackage{graphicx,xcolor}

\usepackage{pdfpages} 
\usepackage{pgffor} 

\makeatletter
\AtBeginDocument{\let\LS@rot\@undefined}
\makeatother

\def\supplementfilename{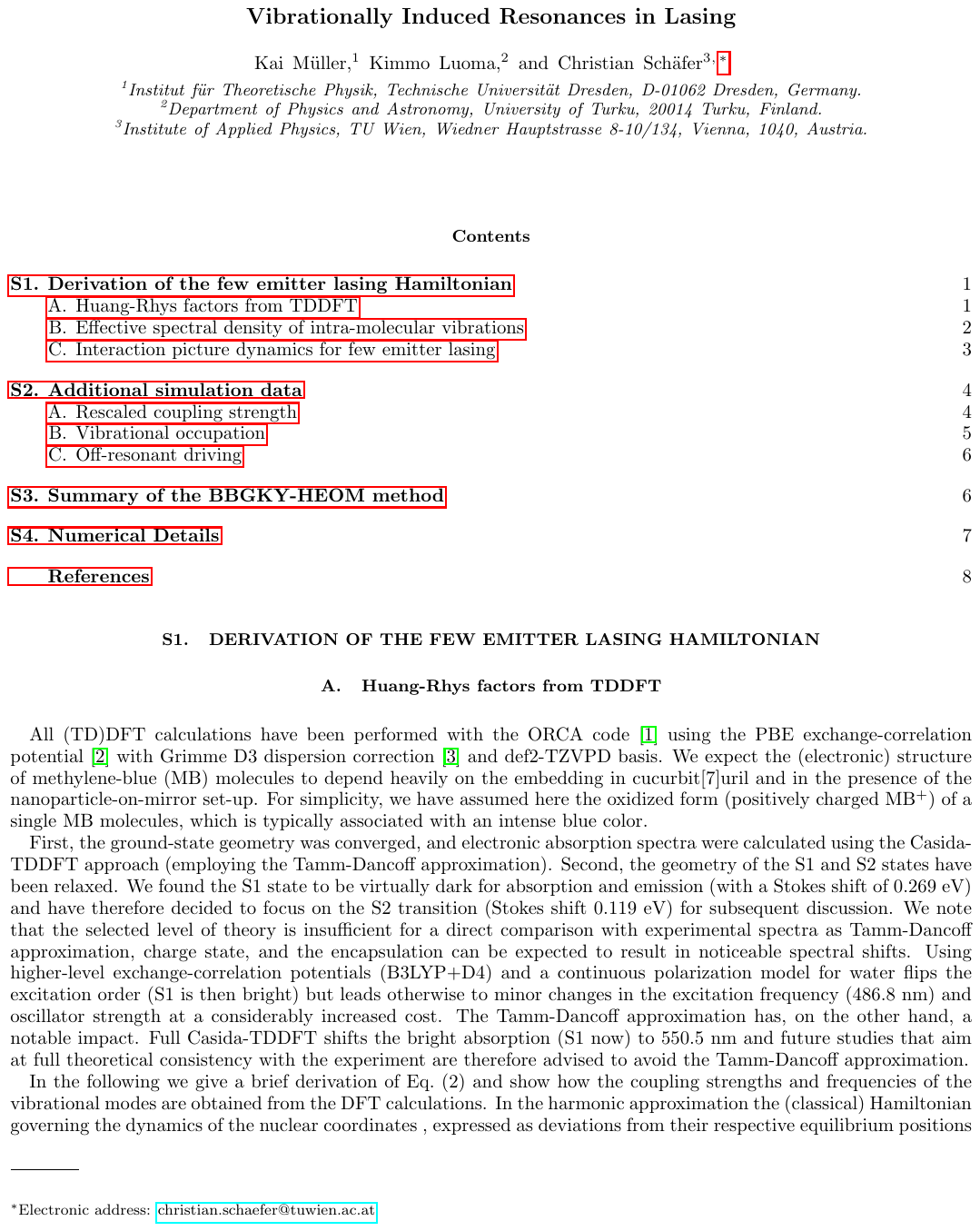}

\pdfximage{\supplementfilename}
\def\numbersupplementpages{\the\pdflastximagepages}

\newif\ifarXiv
\arXivtrue 

\renewcommand{\abs}[1]{\lvert #1\rvert}
\newcommand{\cavg}[1]{\langle #1 \rangle}

\newcommand{\remove}[1]{\textcolor{blue}{\ifmmode\text{\sout{\ensuremath{#1}}}\else\sout{#1}\fi}}
\newcommand{\add}[1]{#1}

\title{Vibrationally Induced Resonances in Lasing}

  \author{Kai M\"uller}
  \affiliation{Institut für Theoretische Physik, Technische Universität Dresden, D-01062 Dresden, Germany.}

  \author{Kimmo Luoma}
  \affiliation{Department of Physics and Astronomy, University of Turku, 20014 Turku, Finland.}

  \author{Christian Sch\"afer}
  \email{christian.schaefer@tuwien.ac.at}
  \affiliation{Institute of Applied Physics, TU Wien, Wiedner Hauptstrasse 8-10/134, Vienna, 1040, Austria.}

\let\oldmaketitle\maketitle
\let\maketitle\relax

\begin{document}

\oldmaketitle
\begin{abstract}
Optical circuits and light sources, such as lasers, undergo continuous miniaturization.
In its extreme, nanolasers might be comprised of only a few molecules confined in plasmonic nanoresonators.
Few-emitter lasers promise low energy requirements and fast responses in a footprint that can be inserted into any device or biological tissue.
Utilizing the recently developed stacked hierarchy approach\cite{article_placeholder}\add{, informed from first principles, we demonstrate the impact of vibrational structure on lasing, using the example of few-molecule lasing in plasmonic cavities}.
Explicitly accounting for the entire vibrational manifold unveils resonances in the laser intensity that depend on the Stokes shift, drive strength, and the number of emitters.
Our work identifies the limits of the omnipresent "incoherent drive"-approximation and paves the way for the understanding of nanolasers at the molecular scale.
\begin{tocentry}
  \includegraphics[width=5.1cm]{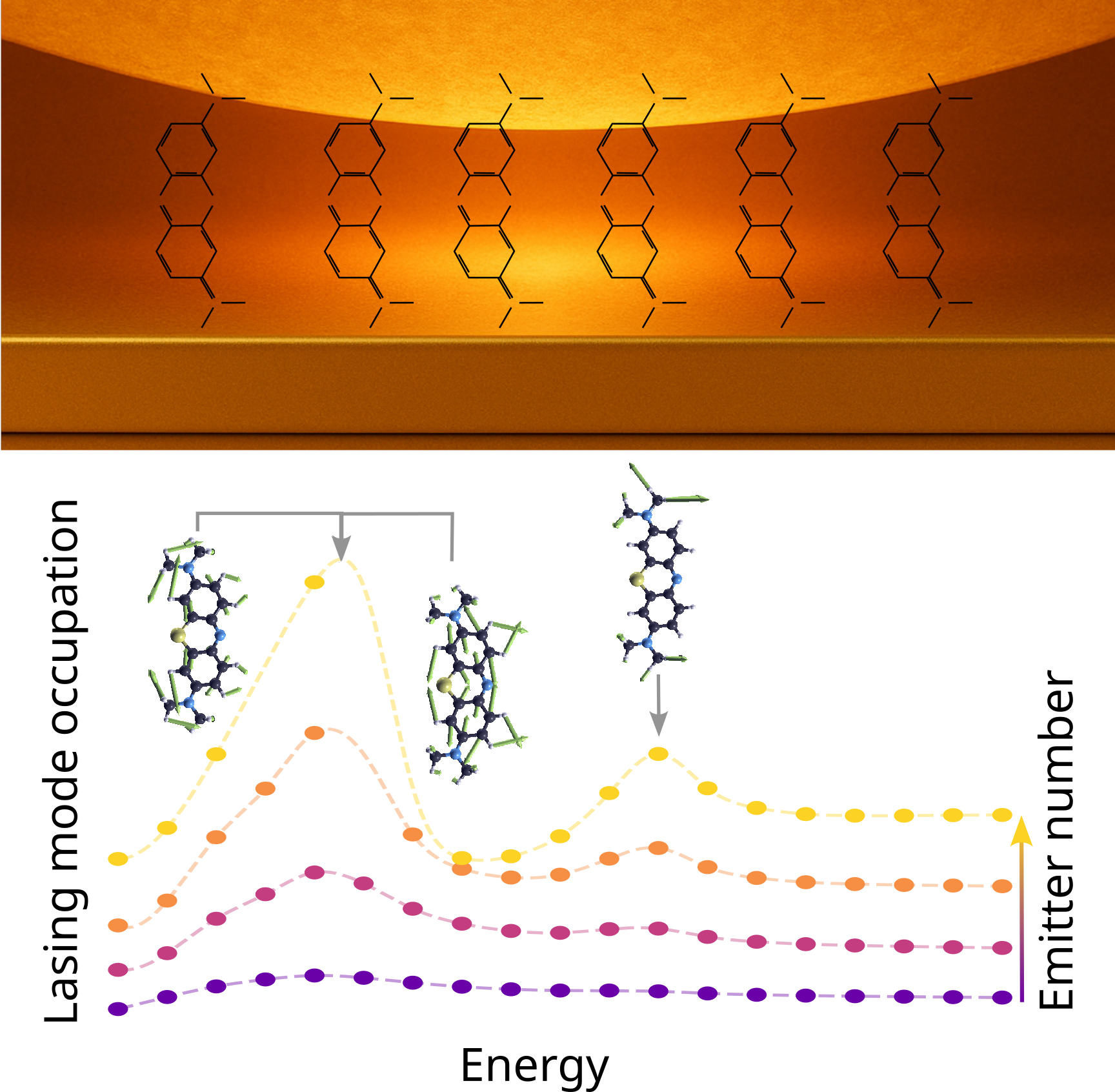}
\end{tocentry}
\end{abstract}

Moors law has dominated the development of electronic components for many decades.
The continuous shrinking of transistors increased the portability and flexibility of our technical devices while capping power usage and boosting performance. 
A similar development can be observed for lasers \cite{ma2019applications,oulton2009plasmon,chow2014emission,reitzenstein2006lasing,strauf2006self,xie2007influence,nomura2010laser}.
Lasing occurs when stimulated emission into a common optical mode overcomes all losses, a process that is commonly linked to a macroscopic gain medium but can be sized down to the nanometer scale.
Nanolaser come in various shapes. 
Whether it be arrays of plasmonic nanoparticles embedded in a gain medium \cite{zhou2013lasing}, or a couple of emitters confined in optical modes with an extension of a few nanometers \cite{FewEmitterLasingKeeling,Ojambati2019Mar}.
Developing a microscopic understanding of such systems can be challenging, especially when the required inversion of occupation necessitates a relaxation via the vibrational manifold.
For a faithful description, the many-body problem of multiple emitters has to be combined with the influence of non-Markovian vibrational baths \cite{PhysRevLett.121.227401,mesoHOPS, Gera2025Jun}. 
Inside a cavity, this leads to a combination of large Hilbert spaces, all-to-all interactions, and non-Markovian evolution that provides a significant barrier for most methods and encourages the use of simplified or effective models.
Common approximations include the Holstein-Tavis-Cummings model \cite{Zeb2018Jan,Herrera2018Jan,holzinger2022cooperative,kirton2013nonequilibrium} where the vibrational spectrum is replaced by a single damped mode, or the combination of coherent drive and vibrational relaxation into an effective incoherent drive \cite{quantumOpticsBook, FewEmitterLasingKeeling,bychek2025nanoscale} assuming fast decay via a structureless vibrational bath.

In this Letter, we provide a detailed theoretical discussion of molecular few-emitter lasing that is largely informed from first principles and accounts for the coherent excitation, inversion, and lasing process without any of the above mentioned approximations. 
We achieve this by utilizing the newly developed stacked BBGKY-HEOM hierarchy, which we present in Ref.~\cite{article_placeholder}. 
The incorporation of the \textit{entire} vibrational manifold for each emitter results in noticeable deviations from paradigmatic approximations and unveils resonant enhancements of the lasing process that are absent in a description that utilizes the widely used approximation of incoherent drives\cite{quantumOpticsBook}.

\begin{figure}
    \centering
    \includegraphics[width=\columnwidth]{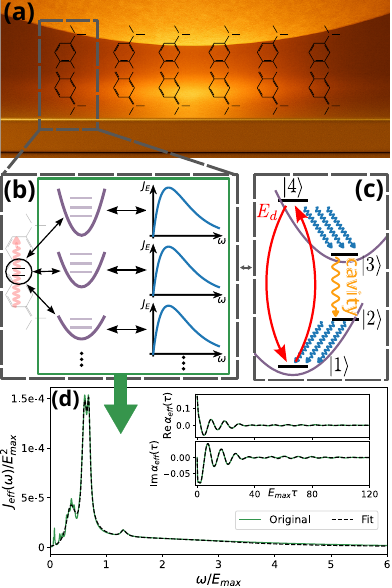}
    \caption{\textbf{Overview of the few emitter lasing system:}(a) Sketch of the plasmonic nanocavity, comprised of a gold nanoparticle above a gold surface. (b) We restrict our model of the electronic states to the first bright transition, leading to a two level system. However, we take the full spectrum of vibrational modes into account, which we approximate as a large set of harmonic oscillators (purple). Each oscillator in turn couples to a continuum of vibrational modes that we model with an ohmic spectral density (blue). All the vibrations can be combined into a single effective bath with the spectral density $J_{eff}(\omega)$ shown in (d) (green solid line). The corresponding bath correlation function shown in the inset is then fitted with exponentials (black dashed line). Graphic (c) shows a sketch of the lasing mechanism, where the states $\ket{2},\,\ket{3}$ are distinguished from the states $\ket{1}, \ket{4}$ by a different vibrational state. Purple parabolas in the background indicate the shifted harmonic potential surfaces.
    }
    \label{fig:sketchMoleculeNanocavity}
\end{figure}

\textit{Molecular lasing informed from first principles --}
We postulate the existence of vibrationally induced resonances in few-emitter lasing, motivated by a recent experimental realization. The underlying physics, as well as the resulting observable signatures, are transferable to a wide range of molecular gain materials.
Figure~\ref{fig:sketchMoleculeNanocavity}(a) illustrates the system under investigation: a small ensemble ($N \approx 2$–$20$) of methylene blue (MB) molecules embedded in a plasmonic nanocavity formed by a gold nanoparticle positioned above a metallic mirror \cite{FewEmitterLasingKeeling,Ojambati2019Mar}.

The locally enhanced fields in the plasmonic nanocavity are dominated by dipolar Mie modes with longitudinal character.
A simplified description of those modes as purely bilinear is reasonable as long as the molecular electronic structure does not influence the plasmonic excitation.\footnote{We refer the interested reader to the ever increasing literature on quadratic corrections for transversal fields \cite{schafer2019relevance,de2025there} and point out that also longitudinal plasmonic excitations will be screened, thus renormalized, when the surface of the nanoplasmonic particle is considerably occupied with molecules \cite{kuisma2022ultrastrong}.}
We therefore start from a Hamiltonian of the form 
\begin{equation}\label{eq:startingH}
    H = \sum_{i=1}^N H_{mol}^{(i)} + H_{drive}+ \sum_{i}g_{cav} (L_{i}a^\dagger + L_i^\dagger a) + \omega_{cav} a^\dagger a,
\end{equation}
where $a$($a^\dagger$) is the annihilation (creation) operator of the cavity mode and $L_i$ is the coupling operator for molecule $i$.
Aggregation, and direct dipole-dipole interaction, is largely avoided in experiment by encapsulating the molecules in dielectric cucurbit[7]uril hosts \cite{FewEmitterLasingKeeling,Ojambati2019Mar}.
In Eq.\eqref{eq:startingH} we have assumed that all molecules couple equally to the cavity mode. While this is not the case in typical experiments, it simplifies the numerical treatment and we argue later that an inhomogeneous coupling strength will not qualitatively affect our results.
For each individual molecule we restrict the electronic degrees of freedom to the many-body states associated with the first bright electronic transition $\widetilde{\omega}_0$. 
However, we take the full spectrum of vibrational modes with frequencies $\omega^{vib}_\lambda$ and Huang-Rhys factors $S_\lambda$ into account.
For a realistic vibrational spectrum, electronic and vibrational frequencies as well as Huang-Rhys factors are estimated using time-dependent density-functional theory \cite{neese2022software,doi:10.1021/acs.jctc.8b00841,Grimme2010} for the methylene blue molecule used in experiment \cite{FewEmitterLasingKeeling} (see Supplemental Material (SM)). 
In the double-harmonic Franck–Condon approximation for the electron-phonon coupling, each of the 107 vibrational normal modes is described as a harmonic oscillator, where the equilibrium position of the oscillator is shifted by a displacement $\Delta_\lambda = \sqrt{2S_\lambda/\omega_\lambda}$ if the molecule is in the electronically excited state. 
Within this approximation, our Hamiltonian for a single molecule takes the form 
\begin{align}\label{eq:Hmol}
    H_{mol}^{(i)} =& \frac{\widetilde{\omega}_0}{2} \sigma^z_i + \sum_{\lambda=1}^{107} \frac{p_{i,\lambda}^2}{2} + \frac{(\omega_{\lambda}^{vib})^2}{2}\left(q_{i,\lambda} - \Delta_\lambda\sigma_i^+\sigma_i^-\right)^2,\notag\\
    =& \frac{\omega_0}{2} \sigma_i^z - \frac{1}{\sqrt{2}} \sum_{\lambda=1}^{107} {\omega_\lambda^{vib}}\sqrt{S_\lambda} \sigma_i^+\sigma_i^-(b_{i,\lambda}+b_{i,\lambda}^\dagger)\\ 
    +& \sum_{\lambda=1}^{107} \omega_{\lambda}^{vib}b_{i,\lambda}^\dagger b_{i,\lambda} \notag
\end{align}
\begin{figure*}[t]
    \centering
    \includegraphics[width=\textwidth]{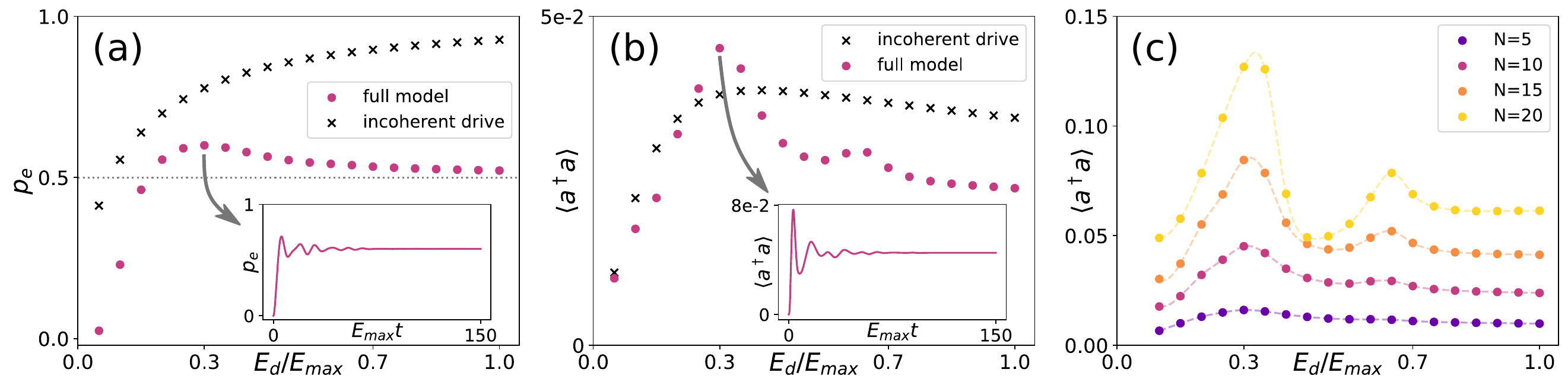}
    \caption{\textbf{Vibrational impact on few-emitter lasing:} We compare the qualitative steady state behavior for increasing the coherent drive in the full model \eqref{eq:H_fewEmitterLasing},\eqref{eq:fewEmitterLasingLocalDiss} against an effective incoherent drive that neglects the vibrational structure \eqref{eq:incoherentLasing}.
    Plots (a) and (b) show the occupation of the excited electronic state, $p_e$, and the cavity mode, $a^{\dagger}a$, respectively, for a plasmonic cavity containing ten molecules. Both plots show maxima arising from resonances with the vibrational spectrum that are missed by an effective treatment with an incoherent drive. Those maxima become more pronounced for increasing $N$ as shown in plot (c).}
    \label{fig:fewEmitterLasing}
\end{figure*}
where $q_{i,\lambda}$, $p_{i,\lambda}$ are the mass renormalized conjugate position and momenta of the $\lambda$-th normal mode of molecule $i$(see SM for a detailed derivation). The operators $b_\lambda$ ($b_\lambda^\dagger$) are the corresponding annihilation (creation) operators, with $b_\lambda=\sqrt{\omega_\lambda/2}(q_\lambda + ip_\lambda/\omega_\lambda)$\add{. Furthermore, we have introduced the vertical Franck-Condon excitation} energy $\omega_0  = \widetilde{\omega}_0 + \sum_\lambda\omega_\lambda^2\Delta_\lambda^2/2$, here $\omega_0=20 857\mathrm{cm}^{-1}$.
Due to the embedding of the MB molecules in the surrounding dielectric host medium \cite{Ojambati2019Mar}, each of these vibrational modes in turn couples to a continuum of environmental modes, which we describe with an ohmic spectral density $J_E(\omega)$ (see Fig.~\ref{fig:sketchMoleculeNanocavity} (b) and SM).
These environmental modes lead to a broadening of the vibrational spectrum and ensure that the vibrational modes quickly relax to the normal (shifted) equilibrium position if the molecule is in the ground (excited) state \cite{Wolfseder1995Mar}. 
Due to the rapid relaxation, the setup resembles a four-level lasing scheme in the sense of Fig.~\ref{fig:sketchMoleculeNanocavity} (c). Coherent excitation to state $\ket{4}$ is followed by a fast relaxation into the new equilibrium position of the vibrational modes $\ket{3}$, 
from which the state decays into $\ket{2}$ via cavity emission. The decay is then again followed by a fast vibrational relaxation into $\ket{1}$. Note that because the state $\ket{2}$ ($\ket{3}$) arises from the shifted equilibrium positions of the vibrational modes, it is generally not orthogonal to $\ket{1}$ ($\ket{4}$).\\

For an exact treatment of this process we proceed by combining all 107 vibrational modes and their environments in an exact way into a single bath with effective spectral density \cite{Roden2012Nov} $J_{eff}(\omega) = \sum_\lambda \abs{g^{eff}_\lambda}^2\delta(\omega - \omega_\lambda)$ shown in Fig.~\ref{fig:sketchMoleculeNanocavity} (d). This process is performed for each molecule and described in detail in the SM. \\
The ($N$) molecules are driven by a coherent field with amplitude $2E_d$ and frequency $\omega_0$ according to $H_{drive} = 2E_d\cos(\omega_0 t)\sum_i\sigma_i^x$.
In an interaction picture that rotates with the drive frequency $\omega_0$ we can perform a rotating wave approximation, assuming that $\omega_0 \gg E_d, g_{cav}$ (see SM).
The resulting Hamiltonian in the interaction picture $H^I$ then takes the form:
\begin{align}\label{eq:H_fewEmitterLasing}
        H^I &= \sum_j\Bigg(E_d\sigma_x^j + \sum_{\lambda}\omega_\lambda^{eff}b_{\lambda,j}^\dagger b_{\lambda,j}\notag\\
        &\quad\quad\quad\quad-\sum_{\lambda} g_\lambda^{eff}\sigma_+^j\sigma_-^j(b_{\lambda,j}^\dagger + b_{\lambda,j}) \Bigg)\notag\\
        & + g_{cav}\sum_j\left(\sigma_j^+ a + \sigma_j^-a^\dagger\right)+(\omega_{cav}-\omega_0)a^\dagger a. 
\end{align}
We include the effects of cavity loss and spontaneous emission by means of the GKSL master equation\cite{Lindblad1976Jun, Gorini1976May} for the total state $\rho_{tot}$
\begin{equation}\label{eq:fewEmitterLasingLocalDiss}
    \begin{split}
        \dot{\rho}_{tot} =& -i[H^I, \rho_{tot}] + \Gamma_{\downarrow}\sum_j\mathcal{D}[\sigma_-^j](\rho_{tot})\\
        &+ \kappa \mathcal{D}[a](\rho_{tot}),
    \end{split}
\end{equation}
with $\mathcal{D}[L](\rho) = 2L\rho L^\dagger - \{L^\dagger L, \rho\}$
To solve Eq.~\eqref{eq:fewEmitterLasingLocalDiss} we utilize the recently developed BBGKY-HEOM approach \cite{article_placeholder}, based on a combination of the Hierarchical Equations of Motion (HEOM) and the Bogoliubov–Born–Green–Kirkwood–Yvon (BBGKY) hierarchy. The method allows for an efficient, approximate description of interacting many-body systems in contact with local or global non-Markovian baths. 
We truncate the BBGKY hierarchy by neglecting \add{beyond Gaussian} three-body correlations.
As a result, the equations of motion of the system become independent of the number of particles $N$. The HEOM then provides an in principle exact treatment of the non-Markovian baths after their bath correlation function has been fitted with exponentials. The details of the method can be found in an accompanying article \cite{article_placeholder}, along with extensive benchmarks for different systems. Here we proceed to show the application of the BBGKY-HEOM method to Eq.~\eqref{eq:fewEmitterLasingLocalDiss} and the physical results.

The fit of the bath correlation function $\alpha_{eff}(\tau) = \int_0^\infty J_{eff}(\omega) e^{-i\omega t}\dd{\omega}$ with a sum of 5 exponentials as $\alpha_{eff}(\tau) \approx \sum_i G_i \exp(-W_i\tau)$ with $G_i,\,W_i\in \mathbb{C}$ is shown in Fig.~\ref{fig:sketchMoleculeNanocavity} (d). 
We treat the cavity as an additional global bath coupling to all molecules. 
Since the bath correlation function of the cavity $\alpha_{cav}(\tau) = \cavg{a(t)a^\dagger(s)} = g_{cav}^2e^{-i\omega_{cav}\tau-\kappa\abs{\tau}}$ is already exponential, no additional fit is required. \add{The full evolution equation for BBGKY-HEOM can be found in the SM.}\\

We now propagate the state of the emitters with different drive strengths $E_d$, up to a maximum drive strength $E_{max}=0.1\omega_0$, and cavity parameters $g_{cav} = 0.2E_{max}$, $\kappa=3.3E_{max}$ until the respective steady states are reached.
The results are shown in Fig.~\ref{fig:fewEmitterLasing} (a), (b), where we show the occupation of the excited electronic state $p_e$ and the cavity in the steady state for ten molecules and different coherent driving strengths $E_d$ (orange circles). Note that the intensity of the laser is proportional to the cavity occupation $I=\kappa\cavg{a^\dagger a}$.
The results are contrasted with a second set of BBGKY-HEOM simulations (black crosses), where the coherent drive and the vibrational bath have been combined into an effective incoherent pumping, such that the molecules evolve according to 
\begin{align}\label{eq:incoherentLasing}
        \dot{\rho}_{incoh} =& -i[H_{incoh}^I, \rho_{incoh}] + \Gamma_{\downarrow}\sum_j\mathcal{D}[\sigma_-^j](\rho_{incoh}) \notag\\
        &+ E_d\sum_j\mathcal{D}[\sigma_+^j](\rho_{incoh}) +\kappa\mathcal{D}[a](\rho_{incoh}),\notag\\
        H_{incoh}^I =& g_{cav}\left(\sum_j\sigma_+^j a + \sum_j\sigma_-^ja^\dagger\right)\notag\\
        &+ (\omega_{cav}-\omega_0)a^\dagger a.
\end{align}
\add{How accurate this approximation is will depend heavily on the coupled bath, i.e., on the molecular and host structure.}
\add{Additionally, note that this comparison is supposed to be of qualitative nature as the different steady-states for a coherent pumping with strength $E_d$ are not directly comparable to those for incoherent pumping with strength $E_d$.}
\add{Instead, we focus our analysis on the qualitative behavior as} \add{the driving strength is increased}.\\
Fig.~\ref{fig:fewEmitterLasing} (a) demonstrates the population inversion required for a lasing transition.
In contrast to the predictions from the incoherent drive, the full model does not indefinitely increase its inversion, but rather reaches a plateau at large driving strengths. There, the
coherent drive leads to a fast polarization that  
dominates the dynamics, resulting in 0.5 inversion.
Interestingly, we find a maximal inversion at $E_d/E_{max} \approx 0.3$, with the corresponding time evolution of $p_e$ shown in the inset.
The steady-state cavity occupation in Fig.~\ref{fig:fewEmitterLasing} (b) features distinct maxima, in addition to a background that is reminiscent of the solution using an incoherent drive. 
These maxima become even more pronounced for larger numbers of molecules $N$, as they are enhanced by the self-amplifying lasing process, see Fig.~\ref{fig:fewEmitterLasing} (c).
\add{This enhancement can be partially understood as collective enhancement $\langle a^\dagger a\rangle \sim N^2$.
However, also the \textit{relative} strength between peaks and background increases with $N$ due to correlated dynamics that extend beyond mean-field theory. We provide a more detailed discussion of this effect in the SM.}\\

\begin{figure}
    \centering
    \includegraphics[width=\columnwidth]{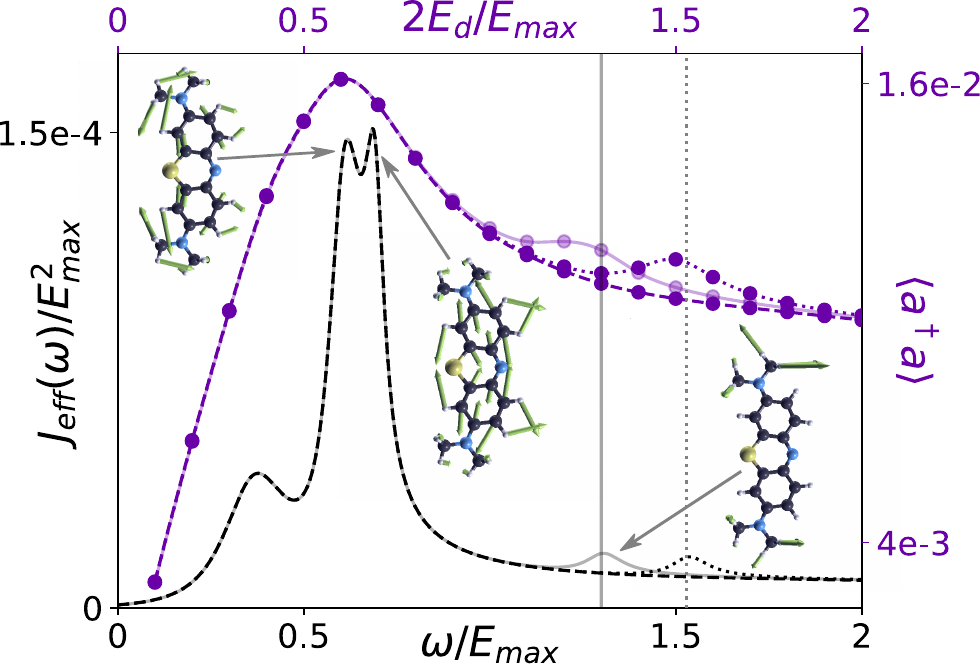}
    \caption{\textbf{Vibrationally induced resonances:} Steady state cavity occupation for $N=5$ (purple) and effective vibrational spectrum are shown on a shared x-axis, with $2E_d = \omega$. Maxima in  $\langle a^\dagger a\rangle$
    line up with peaks in the spectral density. \add{Solid, lighter} 
    lines correspond the the data 
    in Fig.~\ref{fig:fewEmitterLasing}(c) and Fig.~\ref{fig:sketchMoleculeNanocavity}(d) respectively. \add{Intentionally omitting (shifting) a peak in $J_{eff}(\omega)$ also removes (shifts) the corresponding occupation maxima, as evidenced by the darker dashed (dotted) lines.}
    Insets show 
    dominant vibrational modes at the peak positions.
    }
    \label{fig:spectr_comparison}
\end{figure}
To investigate the origin of these maxima, it is instructive to compare their location with the vibrational spectrum $J_{eff}(\omega)$ as shown in Fig.~\ref{fig:spectr_comparison} for the same parameters as in Fig.~\ref{fig:fewEmitterLasing} with $N=5$. We find that the maxima line up with peaks in the effective vibrational spectrum and occur if the drive strength in atomic units $E_d$ is chosen to be half of the peak frequency (solid \add{lighter} line).\\
Artificially removing a peak in the spectral density, e.g. at $\omega/E_{max} \approx 1.3$, we find that also the associated maximum in the cavity occupation vanishes (\add{darker}, dashed lines) and shifting the peak also shifts the corresponding maxima (\add{darker}, dotted lines). 
We thus attribute this effect to resonances between the driving \emph{strength} and the effective vibrational spectrum $J_{eff}$. 
Insets in Fig.~\ref{fig:spectr_comparison} illustrate the dominant vibrational modes.
The occurrence of resonances at $\omega = 2E_d$ can be rationalized from a simplified model of a single dissipation-less vibrational mode for a molecule under \add{resonant} drive $H_{simple} = E_d \sigma_x + g^{vib}\sigma_+\sigma_-(b + b^\dagger) + \omega^{vib}b^\dagger b$. Up to a 90-degree rotation $x\to z,\, z\to -x$ and a shift of the oscillators this model is equivalent to the quantum Rabi model \cite{RabiModel}, which also features a resonance at $\omega=2E_d$. 
\add{In our case this resonance leads to an increased occupation of the vibrational modes, akin to an enhanced occupation of the state $\ket{3}$ in Fig.~\ref{fig:fewEmitterLasing} (c) and thus facilitates the lasing process.}
\add{Intuitively, driving the molecular system results in an AC Stark effect that splits the original states. If the energies of neighboring vibrational states align, which happens precisely at $\omega=2E_d$, energy transfer, vibrational occupation, and thus lasing is maximized.
}
\add{
We provide a more detailed discussion in the SM, Sec.~S2 C.}\\
It is important to note that the resonances are dictated by the peaks in the \textit{effective} vibrational spectrum $J_{eff}$. Thus, their exact position and height are influenced by the spectral density of the environmental modes $J_E(\omega)$ and can in general differ from the frequencies of the isolated modes $\omega_\lambda^{vib}$ in Eq.~\eqref{eq:Hmol}.  
Furthermore, since $J_{eff}$ does not depend on the cavity coupling, we expect that our results still hold for inhomogeneous cavity couplings.
The many-body picture employed in this section provides therefore straightforward access to vibronic coupling and recommends itself as a natural basis to describe few-emitter lasing.\\

\textit{Conclusion --} 
Few-emitter lasers, here molecules in a plasmonic nanocavity, are intricate systems in which optical, thermal, and chemical processes compete.
Fostering this young domain requires a comprehensive understanding at the molecular level, and thus a holistic description.
Instead of combining coherent drive and vibrational relaxation into an effective incoherent drive, we have moved beyond this paradigmatic approximation and accounted for the entire vibrational manifold.
Our approach is largely informed from first principles and based on the recently developed stacked BBGKY-HEOM hierarchy \cite{article_placeholder}.

We observe two major differences from the simplified incoherent model. 
First, maintaining inversion at larger drive strength is challenged by the limited vibrational relaxation speed compared to ever increasing polarization frequencies of the drive-transition.
Second, and especially surprising, we identify clear resonances in the inversion and mode occupation \add{for specific combinations of driving strength and vibrational frequency.} \add{The resonances are induced when AC-Stark shifts energetically align the Franck-Condon transition with neighboring vibrational states -- populating the vibrational mode and enhancing the lasing transition.}
Those resonant features increase with the number of emitter molecules and become substantial in the double-digit domain.
Few-emitter lasing is thus not only more intricate than widely believed, but also offers a path to count the number of coherently coupled molecules in cavity environments.

The strong confinement of molecules in a domain of only a few nanometers suggests that direct dipole-dipole interactions (despite the embedding), multi-mode coupling, vibrational heating \cite{doi:10.1021/acsnano.4c16535}, or even charge transfer between nanocavity and emitter \cite{Fojt2024,HerJueKes23,BroHalNor15}, could further complicate a faithful description.
Our current level of theory therefore provides a realistic, yet simplified description of molecular few-emitter lasing, and our discussion can be transferred to various realizations of a similar design.
Extensions to the theory can be incorporated into the present approach and will be the subject of future studies.

\section*{Acknowledgement}
We thank Walter Strunz and Alexander Eisfeld for insightful discussions.
C.S. acknowledges support from the Swedish Research Council through Grant No. 2016-06059, the FWF cluster of excellence MECS, and funding from the Horizon Europe research and innovation program of the European Union under the Marie Sk{\l}odowska-Curie grant agreement no.\ 101065117.

\section*{Supporting Information}
The Supplemental Information includes:
A derivation of the effective spectral density based on density-functional theory, a comparison for the absorption spectrum to ORCA, the derivation of the Hamiltonian in interaction picture, and numerical details.

\bibliography{library}

\providecommand{\latin}[1]{#1}
\makeatletter
\providecommand{\doi}
  {\begingroup\let\do\@makeother\dospecials
  \catcode`\{=1 \catcode`\}=2 \doi@aux}
\providecommand{\doi@aux}[1]{\endgroup\texttt{#1}}
\makeatother
\providecommand*\mcitethebibliography{\thebibliography}
\csname @ifundefined\endcsname{endmcitethebibliography}
  {\let\endmcitethebibliography\endthebibliography}{}
\begin{mcitethebibliography}{36}
\providecommand*\natexlab[1]{#1}
\providecommand*\mciteSetBstSublistMode[1]{}
\providecommand*\mciteSetBstMaxWidthForm[2]{}
\providecommand*\mciteBstWouldAddEndPuncttrue
  {\def\EndOfBibitem{\unskip.}}
\providecommand*\mciteBstWouldAddEndPunctfalse
  {\let\EndOfBibitem\relax}
\providecommand*\mciteSetBstMidEndSepPunct[3]{}
\providecommand*\mciteSetBstSublistLabelBeginEnd[3]{}
\providecommand*\EndOfBibitem{}
\mciteSetBstSublistMode{f}
\mciteSetBstMaxWidthForm{subitem}{(\alph{mcitesubitemcount})}
\mciteSetBstSublistLabelBeginEnd
  {\mcitemaxwidthsubitemform\space}
  {\relax}
  {\relax}

\bibitem[Müller \latin{et~al.}(2025)Müller, Luoma, and
  Schäfer]{article_placeholder}
Müller,~K.; Luoma,~K.; Schäfer,~C. A Hierarchical Approach to Quantum
  Many-Body Systems in Structured Environments. \textbf{2025}, \relax
\mciteBstWouldAddEndPunctfalse
\mciteSetBstMidEndSepPunct{\mcitedefaultmidpunct}
{}{\mcitedefaultseppunct}\relax
\EndOfBibitem
\bibitem[Ma and Oulton(2019)Ma, and Oulton]{ma2019applications}
Ma,~R.-M.; Oulton,~R.~F. Applications of nanolasers. \emph{Nature
  nanotechnology} \textbf{2019}, \emph{14}, 12--22\relax
\mciteBstWouldAddEndPuncttrue
\mciteSetBstMidEndSepPunct{\mcitedefaultmidpunct}
{\mcitedefaultendpunct}{\mcitedefaultseppunct}\relax
\EndOfBibitem
\bibitem[Oulton \latin{et~al.}(2009)Oulton, Sorger, Zentgraf, Ma, Gladden, Dai,
  Bartal, and Zhang]{oulton2009plasmon}
Oulton,~R.~F.; Sorger,~V.~J.; Zentgraf,~T.; Ma,~R.-M.; Gladden,~C.; Dai,~L.;
  Bartal,~G.; Zhang,~X. Plasmon lasers at deep subwavelength scale.
  \emph{nature} \textbf{2009}, \emph{461}, 629--632\relax
\mciteBstWouldAddEndPuncttrue
\mciteSetBstMidEndSepPunct{\mcitedefaultmidpunct}
{\mcitedefaultendpunct}{\mcitedefaultseppunct}\relax
\EndOfBibitem
\bibitem[Chow \latin{et~al.}(2014)Chow, Jahnke, and Gies]{chow2014emission}
Chow,~W.~W.; Jahnke,~F.; Gies,~C. Emission properties of nanolasers during the
  transition to lasing. \emph{Light: Science \& Applications} \textbf{2014},
  \emph{3}, e201--e201\relax
\mciteBstWouldAddEndPuncttrue
\mciteSetBstMidEndSepPunct{\mcitedefaultmidpunct}
{\mcitedefaultendpunct}{\mcitedefaultseppunct}\relax
\EndOfBibitem
\bibitem[Reitzenstein \latin{et~al.}(2006)Reitzenstein, Bazhenov, Gorbunov,
  Hofmann, M{\"u}nch, L{\"o}ffler, Kamp, Reithmaier, Kulakovskii, and
  Forchel]{reitzenstein2006lasing}
Reitzenstein,~S.; Bazhenov,~A.; Gorbunov,~A.; Hofmann,~C.; M{\"u}nch,~S.;
  L{\"o}ffler,~A.; Kamp,~M.; Reithmaier,~J.; Kulakovskii,~V.; Forchel,~A.
  Lasing in high-Q quantum-dot micropillar cavities. \emph{Applied Physics
  Letters} \textbf{2006}, \emph{89}\relax
\mciteBstWouldAddEndPuncttrue
\mciteSetBstMidEndSepPunct{\mcitedefaultmidpunct}
{\mcitedefaultendpunct}{\mcitedefaultseppunct}\relax
\EndOfBibitem
\bibitem[Strauf \latin{et~al.}(2006)Strauf, Hennessy, Rakher, Choi, Badolato,
  Andreani, Hu, Petroff, and Bouwmeester]{strauf2006self}
Strauf,~S.; Hennessy,~K.; Rakher,~M.; Choi,~Y.-S.; Badolato,~A.; Andreani,~L.;
  Hu,~.~f.~E.; Petroff,~P.; Bouwmeester,~D. Self-tuned quantum dot gain in
  photonic crystal lasers. \emph{Physical review letters} \textbf{2006},
  \emph{96}, 127404\relax
\mciteBstWouldAddEndPuncttrue
\mciteSetBstMidEndSepPunct{\mcitedefaultmidpunct}
{\mcitedefaultendpunct}{\mcitedefaultseppunct}\relax
\EndOfBibitem
\bibitem[Xie \latin{et~al.}(2007)Xie, G{\"o}tzinger, Fang, Cao, and
  Solomon]{xie2007influence}
Xie,~Z.; G{\"o}tzinger,~S.; Fang,~W.; Cao,~H.; Solomon,~G.~S. Influence of a
  single quantum dot state on the characteristics of a microdisk laser.
  \emph{Physical review letters} \textbf{2007}, \emph{98}, 117401\relax
\mciteBstWouldAddEndPuncttrue
\mciteSetBstMidEndSepPunct{\mcitedefaultmidpunct}
{\mcitedefaultendpunct}{\mcitedefaultseppunct}\relax
\EndOfBibitem
\bibitem[Nomura \latin{et~al.}(2010)Nomura, Kumagai, Iwamoto, Ota, and
  Arakawa]{nomura2010laser}
Nomura,~M.; Kumagai,~N.; Iwamoto,~S.; Ota,~Y.; Arakawa,~Y. Laser oscillation in
  a strongly coupled single-quantum-dot--nanocavity system. \emph{Nature
  Physics} \textbf{2010}, \emph{6}, 279--283\relax
\mciteBstWouldAddEndPuncttrue
\mciteSetBstMidEndSepPunct{\mcitedefaultmidpunct}
{\mcitedefaultendpunct}{\mcitedefaultseppunct}\relax
\EndOfBibitem
\bibitem[Zhou \latin{et~al.}(2013)Zhou, Dridi, Suh, Kim, Co, Wasielewski,
  Schatz, and Odom]{zhou2013lasing}
Zhou,~W.; Dridi,~M.; Suh,~J.~Y.; Kim,~C.~H.; Co,~D.~T.; Wasielewski,~M.~R.;
  Schatz,~G.~C.; Odom,~T.~W. Lasing action in strongly coupled plasmonic
  nanocavity arrays. \emph{Nature nanotechnology} \textbf{2013}, \emph{8},
  506--511\relax
\mciteBstWouldAddEndPuncttrue
\mciteSetBstMidEndSepPunct{\mcitedefaultmidpunct}
{\mcitedefaultendpunct}{\mcitedefaultseppunct}\relax
\EndOfBibitem
\bibitem[Ojambati \latin{et~al.}(2024)Ojambati,
  Arnard{\ifmmode\acute{o}\else\'{o}\fi}ttir, Lovett, Keeling, and
  Baumberg]{FewEmitterLasingKeeling}
Ojambati,~O.~S.; Arnard{\ifmmode\acute{o}\else\'{o}\fi}ttir,~K.~B.;
  Lovett,~B.~W.; Keeling,~J.; Baumberg,~J.~J. Few-emitter lasing in single
  ultra-small nanocavities. \emph{Nanophotonics} \textbf{2024}, \emph{13},
  2679--2686\relax
\mciteBstWouldAddEndPuncttrue
\mciteSetBstMidEndSepPunct{\mcitedefaultmidpunct}
{\mcitedefaultendpunct}{\mcitedefaultseppunct}\relax
\EndOfBibitem
\bibitem[Ojambati \latin{et~al.}(2019)Ojambati, Chikkaraddy, Deacon, Horton,
  Kos, Turek, Keyser, and Baumberg]{Ojambati2019Mar}
Ojambati,~O.~S.; Chikkaraddy,~R.; Deacon,~W.~D.; Horton,~M.; Kos,~D.;
  Turek,~V.~A.; Keyser,~U.~F.; Baumberg,~J.~J. Quantum electrodynamics at room
  temperature coupling a single vibrating molecule with a plasmonic nanocavity.
  \emph{Nat. Commun.} \textbf{2019}, \emph{10}, 1--7\relax
\mciteBstWouldAddEndPuncttrue
\mciteSetBstMidEndSepPunct{\mcitedefaultmidpunct}
{\mcitedefaultendpunct}{\mcitedefaultseppunct}\relax
\EndOfBibitem
\bibitem[del Pino \latin{et~al.}(2018)del Pino, Schr\"oder, Chin, Feist, and
  Garcia-Vidal]{PhysRevLett.121.227401}
del Pino,~J.; Schr\"oder,~F. A. Y.~N.; Chin,~A.~W.; Feist,~J.;
  Garcia-Vidal,~F.~J. Tensor Network Simulation of Non-Markovian Dynamics in
  Organic Polaritons. \emph{Phys. Rev. Lett.} \textbf{2018}, \emph{121},
  227401\relax
\mciteBstWouldAddEndPuncttrue
\mciteSetBstMidEndSepPunct{\mcitedefaultmidpunct}
{\mcitedefaultendpunct}{\mcitedefaultseppunct}\relax
\EndOfBibitem
\bibitem[Citty \latin{et~al.}(2024)Citty, Lynd, Gera, Varvelo, and
  Raccah]{mesoHOPS}
Citty,~B.; Lynd,~J.~K.; Gera,~T.; Varvelo,~L.; Raccah,~D. I. G.~B. {MesoHOPS:
  Size-invariant scaling calculations of multi-excitation open quantum
  systems}. \emph{J. Chem. Phys.} \textbf{2024}, \emph{160}\relax
\mciteBstWouldAddEndPuncttrue
\mciteSetBstMidEndSepPunct{\mcitedefaultmidpunct}
{\mcitedefaultendpunct}{\mcitedefaultseppunct}\relax
\EndOfBibitem
\bibitem[Gera \latin{et~al.}(2025)Gera, Hartzell, Chen, Eisfeld, and
  Raccah]{Gera2025Jun}
Gera,~T.; Hartzell,~A.; Chen,~L.; Eisfeld,~A.; Raccah,~D. I. G.~B. Formally
  exact fluorescence spectroscopy simulations for mesoscale molecular
  aggregates with {$N^0$} scaling. \emph{J. Chem. Phys.} \textbf{2025},
  \emph{162}\relax
\mciteBstWouldAddEndPuncttrue
\mciteSetBstMidEndSepPunct{\mcitedefaultmidpunct}
{\mcitedefaultendpunct}{\mcitedefaultseppunct}\relax
\EndOfBibitem
\bibitem[Zeb \latin{et~al.}(2018)Zeb, Kirton, and Keeling]{Zeb2018Jan}
Zeb,~M.~A.; Kirton,~P.~G.; Keeling,~J. Exact States and Spectra of
  Vibrationally Dressed Polaritons. \emph{ACS Photonics} \textbf{2018},
  \emph{5}, 249--257\relax
\mciteBstWouldAddEndPuncttrue
\mciteSetBstMidEndSepPunct{\mcitedefaultmidpunct}
{\mcitedefaultendpunct}{\mcitedefaultseppunct}\relax
\EndOfBibitem
\bibitem[Herrera and Spano(2018)Herrera, and Spano]{Herrera2018Jan}
Herrera,~F.; Spano,~F.~C. {Theory of Nanoscale Organic Cavities: The Essential
  Role of Vibration-Photon Dressed States}. \emph{ACS Photonics} \textbf{2018},
  \emph{5}, 65--79\relax
\mciteBstWouldAddEndPuncttrue
\mciteSetBstMidEndSepPunct{\mcitedefaultmidpunct}
{\mcitedefaultendpunct}{\mcitedefaultseppunct}\relax
\EndOfBibitem
\bibitem[Holzinger \latin{et~al.}(2022)Holzinger, Oh, Reitz, Ritsch, and
  Genes]{holzinger2022cooperative}
Holzinger,~R.; Oh,~S.~A.; Reitz,~M.; Ritsch,~H.; Genes,~C. Cooperative
  subwavelength molecular quantum emitter arrays. \emph{Physical Review
  Research} \textbf{2022}, \emph{4}, 033116\relax
\mciteBstWouldAddEndPuncttrue
\mciteSetBstMidEndSepPunct{\mcitedefaultmidpunct}
{\mcitedefaultendpunct}{\mcitedefaultseppunct}\relax
\EndOfBibitem
\bibitem[Kirton and Keeling(2013)Kirton, and Keeling]{kirton2013nonequilibrium}
Kirton,~P.; Keeling,~J. Nonequilibrium model of photon condensation.
  \emph{Physical review letters} \textbf{2013}, \emph{111}, 100404\relax
\mciteBstWouldAddEndPuncttrue
\mciteSetBstMidEndSepPunct{\mcitedefaultmidpunct}
{\mcitedefaultendpunct}{\mcitedefaultseppunct}\relax
\EndOfBibitem
\bibitem[Scully and Zubairy(1997)Scully, and Zubairy]{quantumOpticsBook}
Scully,~M.~O.; Zubairy,~M.~S. \emph{Quantum Optics}; Cambridge University
  Press, 1997\relax
\mciteBstWouldAddEndPuncttrue
\mciteSetBstMidEndSepPunct{\mcitedefaultmidpunct}
{\mcitedefaultendpunct}{\mcitedefaultseppunct}\relax
\EndOfBibitem
\bibitem[Bychek \latin{et~al.}(2025)Bychek, Holzinger, and
  Ritsch]{bychek2025nanoscale}
Bychek,~A.; Holzinger,~R.; Ritsch,~H. Nanoscale Mirrorless Superradiant Lasing.
  \emph{arXiv preprint arXiv:2505.04025} \textbf{2025}, \relax
\mciteBstWouldAddEndPunctfalse
\mciteSetBstMidEndSepPunct{\mcitedefaultmidpunct}
{}{\mcitedefaultseppunct}\relax
\EndOfBibitem
\bibitem[Sch\"afer \latin{et~al.}(2020)Sch\"afer, Ruggenthaler, Rokaj, and
  Rubio]{schafer2019relevance}
Sch\"afer,~C.; Ruggenthaler,~M.; Rokaj,~V.; Rubio,~A. Relevance of the
  quadratic diamagnetic and self-polarization terms in cavity quantum
  electrodynamics. \emph{ACS Photonics} \textbf{2020}, \emph{7}, 975--990\relax
\mciteBstWouldAddEndPuncttrue
\mciteSetBstMidEndSepPunct{\mcitedefaultmidpunct}
{\mcitedefaultendpunct}{\mcitedefaultseppunct}\relax
\EndOfBibitem
\bibitem[de~la Pradilla \latin{et~al.}(2025)de~la Pradilla, Moreno, and
  Feist]{de2025there}
de~la Pradilla,~D.~F.; Moreno,~E.; Feist,~J. There is no ultrastrong coupling
  with photons. \emph{arXiv preprint arXiv:2508.00702} \textbf{2025}, \relax
\mciteBstWouldAddEndPunctfalse
\mciteSetBstMidEndSepPunct{\mcitedefaultmidpunct}
{}{\mcitedefaultseppunct}\relax
\EndOfBibitem
\bibitem[Kuisma \latin{et~al.}(2022)Kuisma, Rousseaux, Czajkowski, Rossi,
  Shegai, Erhart, and Antosiewicz]{kuisma2022ultrastrong}
Kuisma,~M.; Rousseaux,~B.; Czajkowski,~K.~M.; Rossi,~T.~P.; Shegai,~T.;
  Erhart,~P.; Antosiewicz,~T.~J. Ultrastrong coupling of a single molecule to a
  plasmonic nanocavity: a first-principles study. \emph{ACS photonics}
  \textbf{2022}, \emph{9}, 1065--1077\relax
\mciteBstWouldAddEndPuncttrue
\mciteSetBstMidEndSepPunct{\mcitedefaultmidpunct}
{\mcitedefaultendpunct}{\mcitedefaultseppunct}\relax
\EndOfBibitem
\bibitem[Neese(2022)]{neese2022software}
Neese,~F. Software update: The ORCA program system—Version 5.0. \emph{Wiley
  Interdisciplinary Reviews: Computational Molecular Science} \textbf{2022},
  \emph{12}, e1606\relax
\mciteBstWouldAddEndPuncttrue
\mciteSetBstMidEndSepPunct{\mcitedefaultmidpunct}
{\mcitedefaultendpunct}{\mcitedefaultseppunct}\relax
\EndOfBibitem
\bibitem[de~Souza \latin{et~al.}(2019)de~Souza, Farias, Neese, and
  Izsák]{doi:10.1021/acs.jctc.8b00841}
de~Souza,~B.; Farias,~G.; Neese,~F.; Izsák,~R. Predicting Phosphorescence
  Rates of Light Organic Molecules Using Time-Dependent Density Functional
  Theory and the Path Integral Approach to Dynamics. \emph{Journal of Chemical
  Theory and Computation} \textbf{2019}, \emph{15}, 1896--1904, PMID:
  30721046\relax
\mciteBstWouldAddEndPuncttrue
\mciteSetBstMidEndSepPunct{\mcitedefaultmidpunct}
{\mcitedefaultendpunct}{\mcitedefaultseppunct}\relax
\EndOfBibitem
\bibitem[Grimme \latin{et~al.}(2010)Grimme, Antony, Ehrlich, and
  Krieg]{Grimme2010}
Grimme,~S.; Antony,~J.; Ehrlich,~S.; Krieg,~H. A consistent and accurate ab
  initio parametrization of density functional dispersion correction (DFT-D)
  for the 94 elements H-Pu. \emph{The Journal of Chemical Physics}
  \textbf{2010}, \emph{132}, 154104\relax
\mciteBstWouldAddEndPuncttrue
\mciteSetBstMidEndSepPunct{\mcitedefaultmidpunct}
{\mcitedefaultendpunct}{\mcitedefaultseppunct}\relax
\EndOfBibitem
\bibitem[Wolfseder and Domcke(1995)Wolfseder, and Domcke]{Wolfseder1995Mar}
Wolfseder,~B.; Domcke,~W. {Multi-mode vibronic coupling with dissipation.
  Application of the Monte Carlo wavefunction propagation method}. \emph{Chem.
  Phys. Lett.} \textbf{1995}, \emph{235}, 370--376\relax
\mciteBstWouldAddEndPuncttrue
\mciteSetBstMidEndSepPunct{\mcitedefaultmidpunct}
{\mcitedefaultendpunct}{\mcitedefaultseppunct}\relax
\EndOfBibitem
\bibitem[Roden \latin{et~al.}(2012)Roden, Strunz, Whaley, and
  Eisfeld]{Roden2012Nov}
Roden,~J.; Strunz,~W.~T.; Whaley,~K.~B.; Eisfeld,~A. Accounting for
  intra-molecular vibrational modes in open quantum system description of
  molecular systems. \emph{J. Chem. Phys.} \textbf{2012}, \emph{137},
  204110\relax
\mciteBstWouldAddEndPuncttrue
\mciteSetBstMidEndSepPunct{\mcitedefaultmidpunct}
{\mcitedefaultendpunct}{\mcitedefaultseppunct}\relax
\EndOfBibitem
\bibitem[Lindblad(1976)]{Lindblad1976Jun}
Lindblad,~G. {On the generators of quantum dynamical semigroups}. \emph{Commun.
  Math. Phys.} \textbf{1976}, \emph{48}, 119--130\relax
\mciteBstWouldAddEndPuncttrue
\mciteSetBstMidEndSepPunct{\mcitedefaultmidpunct}
{\mcitedefaultendpunct}{\mcitedefaultseppunct}\relax
\EndOfBibitem
\bibitem[Gorini \latin{et~al.}(1976)Gorini, Kossakowski, and
  Sudarshan]{Gorini1976May}
Gorini,~V.; Kossakowski,~A.; Sudarshan,~E. C.~G. {Completely positive dynamical
  semigroups of N{-}level systems}. \emph{J. Math. Phys.} \textbf{1976},
  \emph{17}, 821--825\relax
\mciteBstWouldAddEndPuncttrue
\mciteSetBstMidEndSepPunct{\mcitedefaultmidpunct}
{\mcitedefaultendpunct}{\mcitedefaultseppunct}\relax
\EndOfBibitem
\bibitem[Xie \latin{et~al.}(2017)Xie, Zhong, Batchelor, and Lee]{RabiModel}
Xie,~Q.; Zhong,~H.; Batchelor,~M.~T.; Lee,~C. {The quantum Rabi model: solution
  and dynamics}. \emph{J. Phys. A: Math. Theor.} \textbf{2017}, \emph{50},
  113001\relax
\mciteBstWouldAddEndPuncttrue
\mciteSetBstMidEndSepPunct{\mcitedefaultmidpunct}
{\mcitedefaultendpunct}{\mcitedefaultseppunct}\relax
\EndOfBibitem
\bibitem[Jakob \latin{et~al.}(2025)Jakob, Juan-Delgado, Mueller, Hu, Arul,
  Boto, Esteban, Aizpurua, and Baumberg]{doi:10.1021/acsnano.4c16535}
Jakob,~L.~A.; Juan-Delgado,~A.; Mueller,~N.~S.; Hu,~S.; Arul,~R.; Boto,~R.~A.;
  Esteban,~R.; Aizpurua,~J.; Baumberg,~J.~J. Optomechanical Pumping of
  Collective Molecular Vibrations in Plasmonic Nanocavities. \emph{ACS Nano}
  \textbf{2025}, \emph{19}, 10977--10988, PMID: 40085022\relax
\mciteBstWouldAddEndPuncttrue
\mciteSetBstMidEndSepPunct{\mcitedefaultmidpunct}
{\mcitedefaultendpunct}{\mcitedefaultseppunct}\relax
\EndOfBibitem
\bibitem[Fojt \latin{et~al.}(2024)Fojt, Erhart, and Schäfer]{Fojt2024}
Fojt,~J.; Erhart,~P.; Schäfer,~C. Controlling Plasmonic Catalysis via Strong
  Coupling with Electromagnetic Resonators. \emph{Nano Letters} \textbf{2024},
  \emph{24}, 11913--11920, PMID: 39264279\relax
\mciteBstWouldAddEndPuncttrue
\mciteSetBstMidEndSepPunct{\mcitedefaultmidpunct}
{\mcitedefaultendpunct}{\mcitedefaultseppunct}\relax
\EndOfBibitem
\bibitem[Herran \latin{et~al.}(2023)Herran, Juergensen, Kessens, Hoeing,
  K{\"o}ppen, {Sousa-Castillo}, Parak, Lange, Reich, Schulz, and
  Cort{\'e}s]{HerJueKes23}
Herran,~M.; Juergensen,~S.; Kessens,~M.; Hoeing,~D.; K{\"o}ppen,~A.;
  {Sousa-Castillo},~A.; Parak,~W.~J.; Lange,~H.; Reich,~S.; Schulz,~F.
  \latin{et~al.}  Plasmonic Bimetallic Two-Dimensional Supercrystals for {{H2}}
  Generation. \emph{Nat. Catal.} \textbf{2023}, \emph{6}, 1205--1214\relax
\mciteBstWouldAddEndPuncttrue
\mciteSetBstMidEndSepPunct{\mcitedefaultmidpunct}
{\mcitedefaultendpunct}{\mcitedefaultseppunct}\relax
\EndOfBibitem
\bibitem[Brongersma \latin{et~al.}(2015)Brongersma, Halas, and
  Nordlander]{BroHalNor15}
Brongersma,~M.~L.; Halas,~N.~J.; Nordlander,~P. Plasmon-Induced Hot Carrier
  Science and Technology. \emph{Nat. Nanotechnol.} \textbf{2015}, \emph{10},
  25--34\relax
\mciteBstWouldAddEndPuncttrue
\mciteSetBstMidEndSepPunct{\mcitedefaultmidpunct}
{\mcitedefaultendpunct}{\mcitedefaultseppunct}\relax
\EndOfBibitem
\end{mcitethebibliography}

\ifarXiv
    

\section*{Supplementary material}

# Vibrationally Induced Resonances in Lasing

Kai Müller,[1] Kimmo Luoma,[2] and Christian Schäfer[3, ∗]

[1]*Institut für Theoretische Physik, Technische Universität Dresden, D-01062 Dresden, Germany.*
[2]*Department of Physics and Astronomy, University of Turku, 20014 Turku, Finland.*
[3]*Institute of Applied Physics, TU Wien, Wiedner Hauptstrasse 8-10/134, Vienna, 1040, Austria.*

## Contents

## S1. DERIVATION OF THE FEW EMITTER LASING HAMILTONIAN

### A. Huang-Rhys factors from TDDFT

All (TD)DFT calculations have been performed with the ORCA code [1] using the PBE exchange-correlation potential [2] with Grimme D3 dispersion correction [3] and def2-TZVPD basis. We expect the (electronic) structure of methylene-blue (MB) molecules to depend heavily on the embedding in cucurbit[7]uril and in the presence of the nanoparticle-on-mirror set-up. For simplicity, we have assumed here the oxidized form (positively charged MB$^+$) of a single MB molecules, which is typically associated with an intense blue color.

First, the ground-state geometry was converged, and electronic absorption spectra were calculated using the Casida-TDDFT approach (employing the Tamm-Dancoff approximation). Second, the geometry of the S1 and S2 states have been relaxed. We found the S1 state to be virtually dark for absorption and emission (with a Stokes shift of 0.269 eV) and have therefore decided to focus on the S2 transition (Stokes shift 0.119 eV) for subsequent discussion. We note that the selected level of theory is insufficient for a direct comparison with experimental spectra as Tamm-Dancoff approximation, charge state, and the encapsulation can be expected to result in noticeable spectral shifts. Using higher-level exchange-correlation potentials (B3LYP+D4) and a continuous polarization model for water flips the excitation order (S1 is then bright) but leads otherwise to minor changes in the excitation frequency (486.8 nm) and oscillator strength at a considerably increased cost. The Tamm-Dancoff approximation has, on the other hand, a notable impact. Full Casida-TDDFT shifts the bright absorption (S1 now) to 550.5 nm and future studies that aim at full theoretical consistency with the experiment are therefore advised to avoid the Tamm-Dancoff approximation.

In the following we give a brief derivation of Eq. (2) and show how the coupling strengths and frequencies of the vibrational modes are obtained from the DFT calculations. In the harmonic approximation the (classical) Hamiltonian governing the dynamics of the nuclear coordinates , expressed as deviations from their respective equilibrium positions

---

∗Electronic address: christian.schaefer@tuwien.ac.at

2can be given in terms of the mass matrix $\mathbf{M}$ and the Hessian Matrix of the potential energy $\mathbf{K}$ (obtained from the DFT calculations) as

$$\begin{aligned} H &= \frac{1}{2}\dot{\eta}^T\mathbf{M}\dot{\eta} + \frac{1}{2}\eta^T\mathbf{K}\eta, \\ &= \frac{1}{2}\dot{\mu}^2 + \frac{1}{2}\mu^T\mathbf{M}^{-1/2}\mathbf{K}\mathbf{M}^{-1/2}\mu, \\ \mu &= \mathbf{M}^{1/2}\eta. \end{aligned} \tag{S1}$$

The mass re-normalized vibrational eigenmodes $u_k$ and the corresponding frequencies $\omega_l$ are now found by means of diagonalization

$$\left[\mathbf{M}^{-1/2}\mathbf{K}\mathbf{M}^{-1/2}\right]\mathbf{u}_k = \omega_k^2 \mathbf{u}_k. \tag{S2}$$

In the electronic groundstate the vibrations are now decoupled in the eigenbasis

$$\begin{aligned} x_k &= \mu \cdot \mathbf{u}_k, \\ H &= \frac{1}{2}\sum_k \dot{x}_k^2 + \omega_k^2 x_k^2 \end{aligned} \tag{S3}$$

and can be canonically quantized

$$\hat{H} = \sum_k \frac{\hat{p}_k^2}{2} + \frac{\omega_k^2}{2}\hat{q}_k^2. \tag{S4}$$

In the double harmonic approximation the dynamics in electronically excited state are governed by the same frequencies but shifted equilibrium positions $\mu \to \mu - \Delta$, where $\Delta$ is the mass weighted nuclear displacement between absorption (at 479.5 nm) and emission (at 502.5 nm) $\Delta = \mathbf{M}^{1/2}(\mathbf{R}_{S0} - \mathbf{R}_{S2})$. Along the $k$-th eigenmode we then obtain a displacement of $\Delta_k = \Delta \cdot \mathbf{u}_k$ in the excited state, leading to the Hamiltonian $H_{mol}$ (Eq. (2) in the letter). The Huang-Rhys factors defined as $S_k = \omega_k \Delta_k^2 / 2$ are thus obtained as

$$S_k = \frac{\omega_k}{2}((\mathbf{R}_{S0} - \mathbf{R}_{S2}) \cdot \sqrt{\mathbf{M}} \cdot \mathbf{u}_k)^2. \tag{S5}$$

### B. Effective spectral density of intra-molecular vibrations

Eq. (2) describes an isolated molecule, where the vibrations form the stick spectrum shown in Fig. S1. To account for the dielectric cage and solvate, in which the molecules are embedded, we couple each molecular vibration to an (accoustic) phonon bath with an ohmic spectral density $J_E(\omega)$. It is important to note that the phonon bath will always relax the vibrational modes to their respective groundstate, which is shifted if the molecule is electronically excited. To ensure this, we describe the coupling between the molecular vibrations and the environmental phonon modes with the coupling operator $b_\lambda - \Delta_\lambda \sigma_+ \sigma_-$ [4]. The total Hamiltonian for a single molecule and the phonon bath with annihilation operators $d_\xi$ now reads

$$\begin{aligned} H_{mol,env} =& \frac{\omega_0}{2}\sigma^z - \frac{1}{\sqrt{2}}\sum_{\lambda=1}^{107}\omega_\lambda^{vib}\sqrt{S_\lambda}\sigma_+\sigma_-(b_\lambda + b_\lambda^\dagger) + \sum_{\lambda=1}^{107}\omega_\lambda^{vib} b_\lambda^\dagger b_\lambda + \sum_{\lambda,\xi}\omega_\xi d_{\lambda,\xi}^\dagger d_{\lambda,\xi} \\ &+ \sum_{\lambda,\xi} g_\xi (b_\lambda - \Delta_\lambda \sigma_+ \sigma_-) d_{\lambda,\xi}^\dagger + \text{h.c.}, \\ J_E(\omega) =& \sum_\xi |g_\xi|^2 \delta(\omega - \omega_\xi), \\ =& A\omega \exp(-\omega/\Lambda)\theta(\omega). \end{aligned} \tag{S6}$$

The choice of $J_E(\omega)$ is motivated by acoustic phonons [5], where the parameter $A$ scales the total coupling strength and thus the broadening of each peak. To achieve a reasonable broadening we choose $A = 0.025$ cm$^{-2}$ and $\Lambda = 3500$ cm$^{-1}$, leading to the spectral density shown in the background of Fig. S1. Various broadenings are tested and we find minor

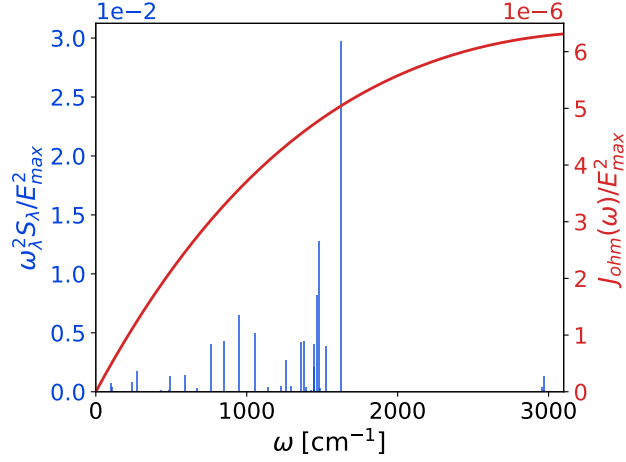

FIG. S1: Squared coupling strengths of the vibrational modes (blue) as obtained from DFT and ohmic spectral density of the phonon environment (red).

influence on the results presented in the main manuscript. The cutoff frequency $\Lambda$ is chosen such that the structure of the resulting spectral density $J_{eff}$ (see below) is not changed upon further increase.

Ref. [6] shows how each of the 107 damped vibrational modes can equally be described by an effective spectral density

$$J_{\lambda,ESM}(\omega) = \frac{\Delta_\lambda^2 A \omega^3 e^{-\omega/\Lambda} \theta(\omega)}{\pi^2 A^2 \omega^2 e^{-2\omega/\Lambda} + \left(\omega - \omega_\lambda + A\Lambda - A\omega e^{-\omega/\Lambda} \operatorname{Ei}(\omega/\Lambda)\right)^2},$$

$$\operatorname{Ei}(x) = -\int_{-x}^{\infty} \frac{e^{-t}}{t}\, dt\,.$$
(S7)

We thus broaden each line in Fig. S2 according to Eq. (S7), which leads us to the effective spectral density seen by the electronic degree of freedom $J_{eff}(\omega)$ shown in Fig. 1(d) and the effective molecular Hamiltonian

$$H_{mol,eff} = \frac{\omega_0}{2}\sigma^z - \sum_\lambda g_\lambda^{eff} \sigma_+ \sigma_- (b_\lambda + b_\lambda^\dagger) + \sum_\lambda \omega_\lambda^{eff} b_\lambda^\dagger b_\lambda,$$

$$J_{eff}(\omega) = \sum_\lambda \left|g_\lambda^{eff}\right|^2 \delta(\omega - \omega_\lambda^{eff}).$$
(S8)

The spectral density $J_{eff}(\omega)$ was used to parameterize our BBGKY-HEOM model by fitting the corresponding correlation function $\alpha_{eff}(\tau) = \int_0^\infty J_{eff}(\omega) e^{-i\omega t}\, d\omega$ with a sum of 5 exponentials. The fit was performed with a simple minimization procedure. In Fig. S2 we compare the resulting absorption spectrum of a single molecule to the path-integral implementation of ORCA [7] and observe qualitative agreement, with some quantitative deviations. A quantitative comparison is complicated by the different treatment of linewidth-broadening in ORCA.

### C. Interaction picture dynamics for few emitter lasing

From the description of a single molecule in Eq. (S8) we now move to multiple molecules that couple identically with strength $g_{cav}$ to a common cavity mode that is in resonance with the electronic transition and are subject to a coherent drive with amplitude $2E_d$ and frequency $\omega_d$

$$H = \sum_j^N H_{mol}^{(j)} + \sum_j^N 2E_d \cos(\omega_d t) \sigma_x^j + \sum_j g_{cav} \sigma_x^j (a + a^\dagger) + \omega_{cav} a^\dagger a.$$
(S9)

Cavity loss and spontaneous decay of the individual molecules are accounted for via a master equation description, such that the total system $\rho_{tot}$ consisting of the molecules and the cavity mode evolves according to

$$\dot{\rho}_{tot} = -i[H, \rho_{tot}] + \kappa \mathcal{D}[a](\rho_{tot}) + \Gamma_\downarrow \sum_i \mathcal{D}[\sigma_-^i](\rho_{tot}).$$
(S10)

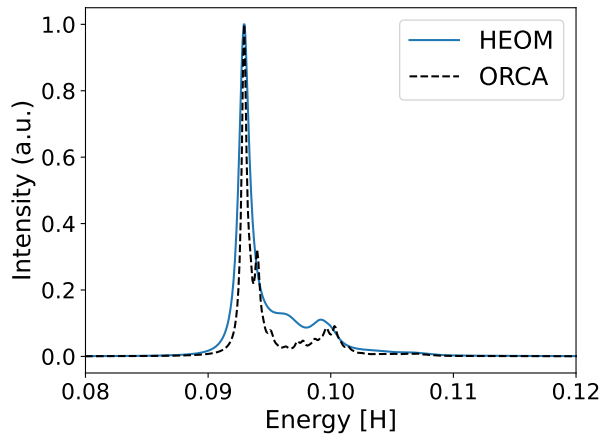

FIG. S2: Absorption spectrum of the molecule based on the Hamiltonian (S8) (blue) compared to the absorption spectrum predicted by ORCA (dashed black). A quantitative comparison is complicated by the different treatment of linewidth-broadening in ORCA.

We now go into an interaction picture $\rho_{tot}^I = U_I \rho U_I^\dagger$ with

$$U_I = \exp\left(i\omega_d t \left(\sum_j \sigma_j^z/2 + a^\dagger a\right)\right), \tag{S11}$$

resulting in the Hamiltonian

$$H^I = \frac{\omega_o - \omega_d}{2}\sum_j \sigma_j^z + E_d \sum_j \sigma_j^x + e^{2i\omega_d t}\sigma_j^+ + e^{-2i\omega_d t}\sigma_j^-$$

$$+ \sum_j \left(-\sum_\lambda g_\lambda^{eff} \sigma_+^j \sigma_-^j (b_{\lambda,j}^\dagger + b_{\lambda,j}) + \sum_\lambda \omega_\lambda^{eff} b_{\lambda,j}^\dagger b_{\lambda,j}\right)$$

$$+ g_{cav}\left(\sum_j \sigma_j^+ a + \sum_j \sigma_j^- a^\dagger + \sum_j e^{2i\omega_d t}\sigma_j^+ a^\dagger + \sum_j e^{-2i\omega_d t}\sigma_j^- a\right)$$

$$+ (\omega_{cav} - \omega_d)a^\dagger a.$$

As long as the driving frequency $\omega_d$ is the fastest timescale in the interaction picture Hamiltonian, we can safely discard counter-rotating terms $\propto e^{\pm 2i\omega_d t}$ that oscillate too fast to influence the dynamics of the system. This in particular implies a drive which is near resonant with both the electronic transition as well as the cavity frequency , sucht that $|\omega_0 - \omega_d|, |\omega_{cav} - \omega_d| \ll \omega_d$. For a resonant drive $\omega_d = \omega_0$ we end up at Eq. (3) of the main letter. Additional data in for off-resonant driving is shown in Sec. S2 C.

## S2. ADDITIONAL SIMULATION DATA

In the following, we present simulation data for some additional parameters which were not discussed in the main text.

### A. Rescaled coupling strength

The first simulation concerns the enhanced lasing intensity for larger $N$, shown in Fig. 2 of the main text. An interesting question is whether this enhancement can be explained by the collectively increased coupling strength $g \to g_{col} = \sqrt{N}g$ typical for the Tavis-Cummings and related models. To answer it we compare the $N = 10, 20$ data

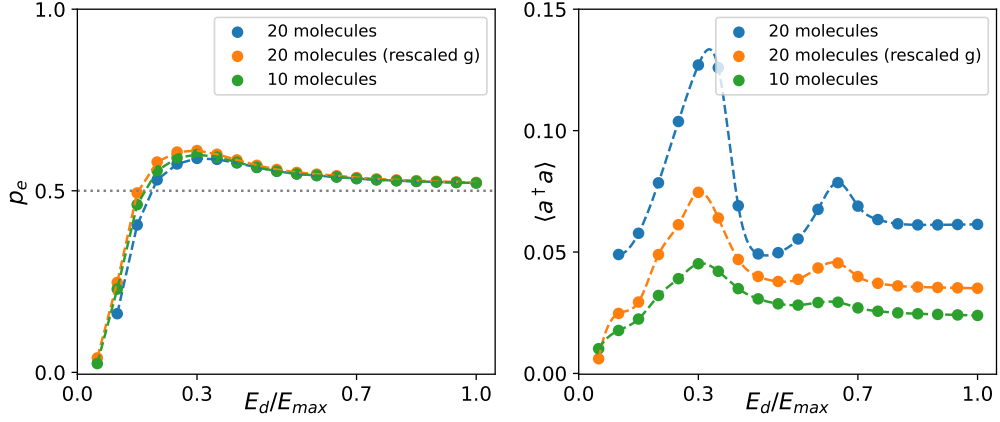

FIG. S3: Electronic inversion and cavity occupation for $N = 10, 20$ molecules from Fig. 2 of the main text compared to a new simulation with $N = 20$ and a rescaled cavity coupling strength $g_{rescaled} = g_{cav}/\sqrt{2}$. Due to the rescaling the collective coupling strength $g_{cav}/\sqrt{N}$ of this new simulation is comparable to the collective coupling strength for $N = 10$ and the original parameters. Nevertheless, the resulting cavity occupation is increased. Dashed lines serve purely as a guide to the eye.

shown in Fig. 2 of the main text with an additional simulation for $N = 20$ molecules, were the coupling strength has been rescaled $g_{rescaled} = g/\sqrt{2}$. Therefore, the collective coupling strength in this new simulation is identical to the collective coupling strength of $N = 10$ molecules for the original parameters (coupling strength $g = 0.2 E_{max}$). The comparison is shown in Fig. S3 and shows the increase in intensity if the collective coupling $g_{cav}\sqrt{N}$ instead of the individual coupling is kept constant.

### B. Vibrational occupation

In addition to the data provided in the main text we discuss the occupation of the vibrational modes for $N = 10$ and otherwise identical parameters to the main text. While we can not resolve the individual occupation of each of the 107 vibrational modes due to the use of the effective spectral density $J_{eff}(\omega)$ and HEOM, we are able to reconstruct the occupation in different regions of the spectrum from the HEOM auxiliary states. Below we distinguish between the occupation of the "first peak" corresponding to the region from $\omega = 0$ to $\omega \approx 1 E_{max}$ and the "second peak", corresponding to the small maximum in $J_{eff}$ at $\omega \approx 1.3 E_{max}$. Fig. S4 compares their occupation for different driving strengths (black curve). The red curve shows the cavity occupation in comparison, which clearly establishes a correlation with the vibrational occupation. The maxima in the vibrational occupation can be explained with the help of $H_{simple} = E_d \sigma_x + g^{vib}\sigma_+ \sigma_-(b + b^\dagger) + \omega^{vib} b^\dagger b$ briefly discussed in the main text. In a rotating frame with respect to the drive $\left|\widetilde{\psi}\right\rangle = \exp\left(i E_d \sigma_x + 2 E_d b^\dagger b\right)|\psi\rangle$ we find

$$\begin{aligned}
\widetilde{H}_{simple} =& (\omega^{vib} - 2E_d) b^\dagger b + g^{vib}(\sigma_{+x} b + \sigma_{-x} b^\dagger) + \frac{g^{vib}}{2}\left(b e^{-2iE_d t} + b^\dagger e^{2iE_d t}\right) \\
& + \frac{g^{vib}}{2}\left(e^{-4iE_d t} b^\dagger \sigma_{+x} + e^{4iE_d t} b \sigma_{-x}\right), \\
\sigma_{+x} =& \frac{\sigma_z - i\sigma_y}{2}, \\
\sigma_{-x} =& \frac{\sigma_z + i\sigma_y}{2}.
\end{aligned} \quad (S12)$$

Neglecting the rotating terms (valid for $\omega^{vib} \approx 2E_d$), clearly show how the interaction becomes resonant at $\omega_{vib} = 2E_d$, as all other terms vanish. We expect that this increases the population in state 3, i.e., the upper state of the lasing transition, which enhances the lasing and thus populates the cavity mode (magenta curves). A more detailed analytical derivation or studies of suitable minimal models that show this effect is an exiting prospect for future studies.

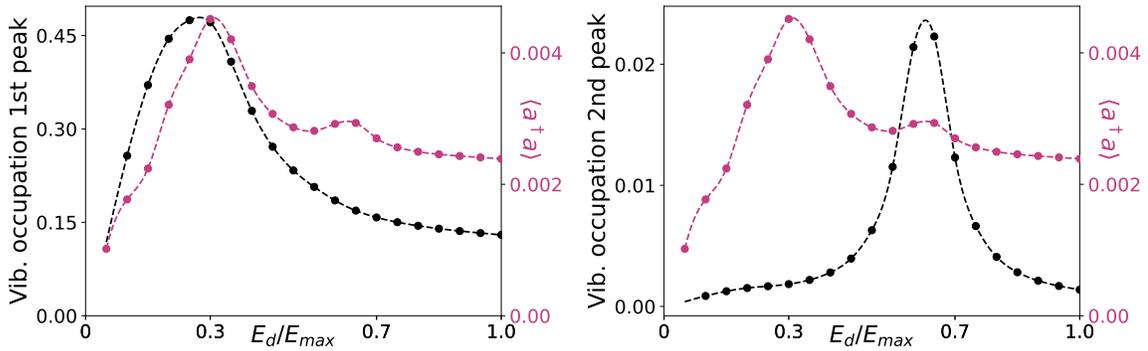

FIG. S4: Vibrational occupation (black dots) over driving strength in different regions of the vibrational spectrum (see text) and $N = 10$. For reference we also show the cavity occupation for the same drive strengths, which highlights a correlation between the cavity and vibrational occupations. Dashed lines serve purely as a guide to the eye.

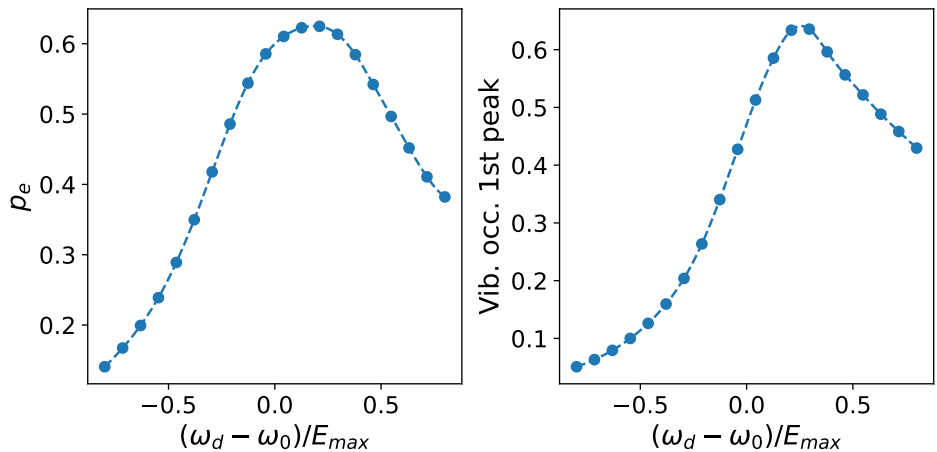

FIG. S5: Inversion and vibrational occupation for different detunings of the drive. Dashed lines serve merely as aguide to the eye.

## C. Off-resonant driving

In the main text we focus on the qualitative features that arise upon changing the driving strength at a fixed frequency, which corresponds to the maximum of the molecular absorption spectrum (without cavity). Fig. S5 shows the behavior of the inversion as well as the vibrational occupation upon varying the drive frequency at a fixed strength $E_d = 0.3 E_{max}$. The maximal values are reached close to, but not exactly at $\omega_0$.

## S3. SUMMARY OF THE BBGKY-HEOM METHOD

In the following we give a brief summary of the BBGKY-HEOM method that was used to obtain the reduced dynamics under Eq. (S12). A detailed derivation of the method including several benchmarks can be found in Ref. [8]. The method is based on a combination of the Bogoliubov-Born-Green-Kirkwood-Yvon (BBGKY) hierarchy and the Hierarchical Equations Of Motion (HEOM) and it allows an approximate treatment of many-body systems in contact with both global and local non-Markovian environments. In the present case we treat the electronic states of the molecules as the system and the damped cavity mode as a global bath (meaning it couples equally to every molecule). In addition, the vibrational manifolds form local non-Markovian baths, meaning they couple to only one molecule and there is one such bath per molecule. The core approximation to make the solution of the model tractable is

introduced by the BBGKY hierarchy and amounts to a Gaussian approximation of three-particle correlations (and higher). Essentially, we approximate the density matrix of three molecules $\rho_{123}$ in terms of the two-molecule density $\rho_{12}$ and the one molecule density matrix $\rho_1$ as

$$\rho_{123} \approx \rho_1 \rho_{23} + \rho_{12}\rho_3 + \rho_{13}\rho_2 - 2\rho_1\rho_2\rho_3, \tag{S13}$$

Due to the particle exchange symmetry of our system the reduced one and two particle density matrices $\rho_k$ and $\rho_{ij}$ ($i,j,k \in [1,2,3]$) correspond to identical operators, that do however act on different particles. The approximation (S13) allows us to derive a closed equation for the reduced state of two molecules $F_{12}^{(\mathbf{0},\mathbf{0})}(t) = N(N-1)\rho_{12}(t)$, where the normalization is by convention. Including the HEOM auxiliary states $\rho^{(\mathbf{n},\mathbf{m})}$, labeled by the vector indices $\mathbf{n}, \mathbf{m}$ the resulting evolution equation reads

$$\begin{aligned}
\dot{F}_{12}^{(\mathbf{n},\mathbf{m})} =& -i[H_{12}, F_{12}^{(\mathbf{n},\mathbf{m})}] - (\mathbf{w}\cdot\mathbf{n} + \mathbf{w}^*\cdot\mathbf{m})F_{12}^{(\mathbf{n},\mathbf{m})} + g_{cav}\left(n_k(L_{g1}+L_{g2})F_{12}^{(\mathbf{n}-\mathbf{e}_k,\mathbf{m})} + n_k\,\mathrm{tr}_3(L_{g3}F_{123}^{(\mathbf{n}-\mathbf{e}_k,\mathbf{m})})\right) \\
&+ g_{cav}\left(m_k F_{12}^{(\mathbf{n},\mathbf{m}-\mathbf{e}_k)}(L_{g1}^\dagger+L_{g2}^\dagger) + m_k\,\mathrm{tr}_3(L_{g3}^\dagger F_{123}^{(\mathbf{n},\mathbf{m}-\mathbf{e}_k)}) + \left[F_{12}^{(\mathbf{n}+\mathbf{e}_k,\mathbf{m})}, L_{g1}^\dagger+L_{g2}^\dagger\right] + \left[L_{g1}+L_{g2}, F_{12}^{(\mathbf{n},\mathbf{m}+\mathbf{e}_k)}\right]\right) \\
&+ \sum_{k=2}^{2M+1} \sqrt{G_k} n_k L_k F_{12}^{(\mathbf{n}-\mathbf{e}_k,\mathbf{m})} + \sqrt{G_k}^* m_k F_{12}^{(\mathbf{n},\mathbf{m}-\mathbf{e}_k)}L_k^\dagger + \sqrt{G_k}\left[F_{12}^{(\mathbf{n}+\mathbf{e}_k,\mathbf{m})}, L_k^\dagger\right] + \sqrt{G_k}^*\left[L_k, F_{12}^{(\mathbf{n},\mathbf{m}+\mathbf{e}_k)}\right].
\end{aligned} \tag{S14}$$

We have already adjusted Eq. (S14) to our specific application, where $L_{gj} = \sigma_j^-$ ($j=1,2,3$) is the coupling operator to the global bath (here the cavity mode), $L_k = \sigma_1^+\sigma_1^-$ for even $k$ and $L_k = \sigma_2^+\sigma_2^-$ for odd $k$, which represents the coupling to the local vibrational baths of molecule one and two. The vibrational bath is fitted with $M$ exponentials as $\alpha_{eff}(\tau) = \sum_{k=1}^{M} G_{2k+1}e^{-w_{2k+1}\tau}$. Including the hierarchy index the approximation of the three particle matrices in terms of the two particle ones are given as [8]

$$\begin{aligned}
F_{123}^{(\mathbf{n},\mathbf{m})} =& 4\frac{(N-1)(N-2)}{N^3}\,\mathrm{tr}\left(F_1^{(\mathbf{n},\mathbf{m})}\right)F_1 F_2 F_3 + \frac{N-2}{N}\left(F_{12}F_3^{(\mathbf{n},\mathbf{m})} + F_2^{(\mathbf{n},\mathbf{m})}F_{13} + F_1^{(\mathbf{n},\mathbf{m})}F_{23}\right) \\
&+ \frac{N-2}{N}\left(F_1 F_{23}^{(\mathbf{n},\mathbf{m})} + F_{13}^{(\mathbf{n},\mathbf{m})}F_2 + F_{12}^{(\mathbf{n},\mathbf{m})}F_3\right) - \frac{N-2}{N^2}\,\mathrm{tr}\left(F_1^{(\mathbf{n},\mathbf{m})}\right)(F_{12}F_3 + F_{13}F_2 + F_1 F_{23}) \\
&- 2\frac{(N-2)(N-1)}{N^2}\left(F_1^{(\mathbf{n},\mathbf{m})}F_2 F_3 + F_1 F_2^{(\mathbf{n},\mathbf{m})}F_3 + F_1 + F_2 F_3^{(\mathbf{n},\mathbf{m})}\right),
\end{aligned}$$

where $F_j = N\rho_j$.

### S4. NUMERICAL DETAILS

Below we provide additional details on how the simulations shown in Fig. 2 and 3 in the letter were performed. All simulations rely on the fit of the effective bath correlation function shown in Fig. 1(d) in the letter. This fit was performed using five exponential functions and the optimal parameters were obtained by minimizing the $L^1$ distance over the relevant time interval. To ensure the correctness of our algorithm and Eq. (S7) we checked that an explicit implementation of Eq. (S6) with a *single* vibrational mode (taken into the system) coupled to the ohmic environment $J_E$ agreed with the corresponding BBGKY-HEOM implementation of the effective vibrational bath (S8). For the final simulations we used a hierarchy depth of five for the vibrations and three for the cavity mode. We checked for selected parameters that the results don't change significantly (deviation smaller than the linewidth in the plots) upon increasing the hierarchy depth or when using a fit with 6 exponentials.

Finally, one may ask whether the BBGKY framework is even necessary to investigate this system or whether one might just as well use a mean-field approximation. To answer this question, we compare the reduced electronic state of two molecules obtained from Eq. (S14) $F_{12}(t)$ with its factorized approximation $\mathrm{Tr}_2(F_{12})\otimes\mathrm{Tr}_2(F_{12})$. The relative error of this approximation is obtained with the help of the Frobenius norm $\|\cdot\|$ as $\|F_{12}(t) - \mathrm{Tr}_2(F_{12}(t))\otimes\mathrm{Tr}_2(F_{12}(t))\|/\|F_{12}(t)\|$ and shown in Fig. S6 for a driving strength $E_d = 0.65 E_{max}$ and different atom numbers. The relative error exceeds 6% for the data discussed in the main text. This would accumulate during time-evolution under a mean-field approximation, highlighting the importance of taking correlations into account. Furthermore, we find that for the parameters considered in the main text ($g_{cav}$ independent of $N$) the relative error increases with the atom number. Rescaling the coupling strength $g \to g/\sqrt{N}$ restores the usual intuition that in all-to-all interacting models mean-field approximations become accurate for large $N$.

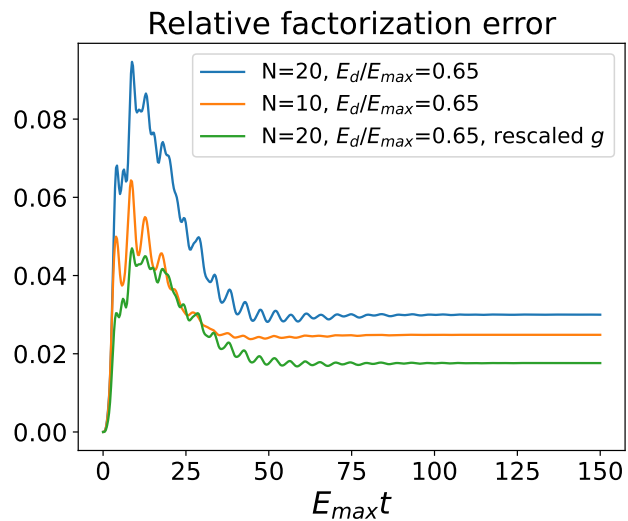

FIG. S6: The relative error of factorising the reduced two-particle density matrix obtained from BBGKY-HEOM as $F_{12}(t) \approx \text{Tr}_2(F_{12}) \otimes \text{Tr}_2(F_{12})$ for $E_d = 0.65 E_{max}$. The relative error is obtained based on the Frobenius norm and shown for different atom numbers. The green curve corresponds to a simulation with $N = 20$ molecules and the same collective coupling strength $\sqrt{N} g_{cav}$ as the orange curve, where $N = 10$ molecules were included.

[1] F. Neese, Wiley Interdisciplinary Reviews: Computational Molecular Science **12**, e1606 (2022).
[2] J. P. Perdew, K. Burke, and Y. Wang, Physical Review B **54**, 16533 (1996), ISSN 1095-3795.
[3] S. Grimme, J. Antony, S. Ehrlich, and H. Krieg, The Journal of Chemical Physics **132**, 154104 (2010), ISSN 0021-9606.
[4] B. Wolfseder and W. Domcke, Chem. Phys. Lett. **235**, 370 (1995), ISSN 0009-2614.
[5] U. Weiss, *Quantum Dissipative Systems (Third Edition)*, Series In Modern Condensed Matter Physics (World Scientific Publishing Company, 2008), ISBN 9789814471855, URL https://books.google.de/books?id=p0jICgAAQBAJ.
[6] J. Roden, W. T. Strunz, K. B. Whaley, and A. Eisfeld, J. Chem. Phys. **137**, 204110 (2012), ISSN 0021-9606.
[7] B. de Souza, G. Farias, F. Neese, and R. Izsák, Journal of Chemical Theory and Computation **15**, 1896 (2019), pMID: 30721046.
[8] K. Müller, K. Luoma, and C. Schäfer, *A hierarchical approach to quantum many-body systems in structured environments* (2025), 2405.05093, URL https://arxiv.org/abs/2405.05093.


\fi

\end{document}